\documentclass{iaus}
\usepackage{graphicx}
\usepackage{epsfig}

\title[Simulator for Microlens Planet Surveys] 
{Simulator for Microlens Planet Surveys}

\author[Sergei I. Ipatov et al.]   
{Sergei I. Ipatov$^1$, Keith Horne$^2$, Khalid A. Alsubai$^3$,
Daniel M.  Bramich$^4$, Martin  Dominik$^2$, Markus P.G. Hundertmark$^2$,
Christine Liebig$^2$, Colin D.B.  Snodgrass$^5$, Rachel A. Street$^6$,
Yiannis  Tsapras$^{6,7}$
}

\affiliation{$^1$Alsubai Est. for Scientific Studies, Doha, Qatar; 
 email: {\tt siipatov@hotmail.com} 

$^2$University of St. Andrews, St.Andrews, Scotland, United Kingdom

$^3$Qatar Foundation, Doha, Qatar

$^4$European Southern Observatory, Garching bei M\"{u}nchen, Germany

$^5$Max Planck Institute for Solar System Research, Katlenburg-Lindau, Germany

$^6$Las Cumbres Observatory Global Telescope Network, Santa Barbara, USA
  
$^7$School of Physics and Astronomy, Queen Mary, University of London,
United Kingdom} 

\pubyear{2013}
\volume{293}  
\pagerange{ -- }
\jname{Formation, detection, and characterization of extrasolar habitable planets}
\editors{N. Haghighipour, ed.} 
\begin{document}

\maketitle

\begin{abstract}
 
We summarize the status of a computer simulator for microlens planet surveys.  
The simulator generates synthetic light curves of microlensing events observed with specified networks of telescopes over specified periods of time. 
Particular attention is paid to models for sky brightness and seeing, calibrated by fitting to data from the OGLE survey and RoboNet observations in 2011. Time intervals during which events are observable are identified by accounting for positions of the Sun and the Moon, and other restrictions on telescope pointing. Simulated observations are then generated for an algorithm that adjusts target priorities in real time with the aim of maximizing planet detection zone area summed over all the available events. 
The exoplanet detection capability of observations 
was compared for several telescopes.
\end{abstract}

\firstsection 

\section{Main features of the simulator and observations used}

Microlensing is unique in its sensitivity to wider-orbit (i.e. cool) planetary-mass bodies, allowing to find planets with masses down to the mass of the Moon. 
   In the present paper, we discuss the simulator that adjusts target priorities in real time with the aim of maximizing planet detection zone area summed over all the available events. 
The observability of a target is limited by its own position on the sky, as well as that of the Sun and the Moon, and telescopes moreover have pointing restrictions. 
Particular attention is paid to  models for sky brightness and seeing.                                                                        
Based on the approach presented by \cite[Horne et al. (2009)]{Horne_etal09}, at each time step for different events we calculate the detection zone area and the exoplanet detection capability of observations.

We considered the following telescopes numbered from $N_t$=1 to $N_t$=13:

1. 2 m FTS - Faulkes Telescope South - Siding Spring Observatory, Australia.

2. 2 m FTN - Faulkes Telescope North - Haleakela, Hawaii, USA.

3. 2 m LT - Liverpool Telescope - La Palma, Canary Islands, Spain.

4. 1.3 m OGLE - The Optical Gravitational Lensing Experiment 
- Las Campanas, 
Chile.

5-7.  Three 1 m CTIO - Cerro Tololo Inter-American Observatory, Chile .

8.  1 m MDO - McDonald observatory - Texas, USA.

9-11. Three 1 m SAAO - South African Astronomical Observatory, South Africa.

12-13.  Two 1 m SSO - Siding Spring Observatory, Australia.

\section{Results}
   
{\underline {\it Sky brightness and seeing.}} 
For studies of sky brightness for FTS, FTN, and LT, we considered those microlensing events 
observed in 2011 for which 
the minimum number of light curve data points 
in .dat files
is greater than 15: FTS -- 39 events; FTN -- 19 events, LT -- 20 events. For OGLE we considered 20 events (110251-110270).
The value of $I_{\rm sky}(0)$ (I-band sky brightness at the zenith) for the model
 was chosen to minimize the sum of squares of  sky brightness residuals (relative to observations)
in the case when the Moon is below the horizon.
Using $\chi^2$ optimization of the straight line fit, we studied the values of $b_{0}$ and $b_1$ in the 
following relationship for  sky brightness: $b=b_{0}+b_1×(a-1)$, where $a$ is airmass.
 For the relationship $b=b_{0i}+b_2×(a-1)$ with one value of $b_2$ and different values of $b_{0i}$  for different events,
we obtained $\Delta b_0 = \max (b_{0i}) - \min (b_{0i})$
to be about 0.7-1.1 mag if we consider observations for the Moon below the horizon.
The values of
$\Delta b_0$ characterize  the difference in sky brightness due to surrounding stars near different events.
The values of $I_{\rm sky}(0)$ presented in Table 1 were obtained 
for an I-band extinction coefficient $e_I=0.05$ mag/airmass.
 For $e_I$ equal to 0 and 0.1, the values of $I_{\rm sky}(0)$ differed by less 
than 0.1 mag from the values at $e_I=0.05$.
The table also presents the values of $ b_0$, $b_1$, $b_2$, $\min (b_{0i})$, and $\max (b_{0i})$.
In Table 1 one can also find the values of $s_0$, $s_1$, and $\sigma_s$ obtained by $\chi^2$ optimization of the straight line fit
for seeing $s$ (FWHM in arcsec) vs. airmass $a \approx \sec{z}$ 
(where $z$ is the zenith distance): $s=s_0+s_1\,(a-1)$ ($\chi^2=\sum [(s_i^\circ -s_1 (a_i - 1)-s_0)/\sigma_s]^2$, 
where $\sigma_s^2$  is variance, the sum is for considered observations, 
$s_i^\circ$ are the known values of $s$ at $a = a_i$).

\begin{table}
  \begin{center}
  \caption{The values $I_{\rm sky}(0)$ of sky brightness at zenith (I magnitude per square arcsec) for the model used, and the
coefficients $b_0$, $b_1$, $ \min (b_{0i})$, $\max (b_{0i})$, $b_2$ characterizing the sky brightness
for the Moon below the horizon.
Coefficients $s_0$ and $s_1$ characterize the dependence of  seeing $s$ (FWHM in arcsec) 
on airmass $a$ ($\chi^2$ optimization: $s=s_0+s_1 (a-1)$).}
  \label{tab2}
 {\scriptsize
  \begin{tabular}{|l|c|c|c|c|}\hline 
{\bf telescope} & {\bf FTS} & {\bf FTN} & {\bf LT} & {\bf OGLE} \\ \hline
$I_{\rm sky}(0)$  & 19.0	& 18.7	& 19.6	& 18.1 \\  \hline
$ b_0          $  & 18.8	& 18.3	& 19.0	& 18.0 \\  \hline
$ b_1          $  & -0.142	& -0.132& -0.110& -0.220 \\  \hline
$ \min (b_{0i}), \max (b_{0i})$ &18.2, 19.3  &  17.9, 19.0& 18.7, 19.4& 17.8, 18.5 \\  \hline
$ b_2          $  & -0.206	& -0.183& -0.259& -0.235 \\  \hline
$ s_0$         &	1.33	&0.69&	1.35	&1.33\\  \hline
$ s_1$  	&0.52&	0.21	&0.42&	0.29\\  \hline
$\sigma_s$	&0.37	&0.21	&0.50 &	0.25\\  \hline
  \end{tabular}
  }
 \end{center}
\end{table}

In order to understand the influence of positions of the Moon and the Sun on a typical sky brightness near an event, 
we compared the minimum and maximum values of $b_{0i}$ for (1) all observations, (2) the Moon below the horizon, and (3) the Moon below the horizon and solar elevation $\theta_{\rm Sun}< -18^\circ$. 
The maximum values of $b_{0i}$ (less bright sky) are almost the same
for different positions of the Moon and the Sun. 
The difference in the lower limit of $b_{0i}$ 
is greater and can be up to 1.5 mag. 
At the Moon below the horizon and $\theta_{\rm Sun} < -18^\circ$, the value of  $\sigma_b$ (the square root of variance) 
can be smaller than that for all observations by a factor of 3 (e.g., it is 0.11 instead of 0.36 for FTS).                                                                                                

{\underline {\it The influence of solar elevation on sky brightness.}} 
Most of 
sky brightness residuals relative to the best fit model
for each event
are in a small range (-0.4 to 0.4 mag) even for all Moon and Sun positions; 
for the Moon below the horizon, there are many values in the range [-0.2, 0.2]; 
greater values of residuals are for a small number of observations. 
   The range of sky brightness residuals for the Moon below the horizon and 
$\theta_{\rm Sun}<-18^\circ$ (for example, [-0.42, 0.87] for FTS) is smaller by a factor of several than 
that for all positions of the Moon and the Sun. 
   For considered observations, 
the lower limit of residuals was greater than $-1$ mag only 
when both the Moon was below the horizon and $\theta_{\rm Sun} <-18^\circ$.                 
                                                      
The influence of solar elevation on sky brightness began to play a role at  $\theta_{\rm Sun}> -14^\circ$, 
and it was considerable at $\theta_{\rm Sun} > -7^\circ$. For example, if we consider only FTS observations for the Moon below the horizon, then sky brightness residual
$s_{br}$ can be about -3 mag at $\theta_{\rm Sun} \in$  [$-8^\circ$, -7$^\circ$],
$s_{br}>-1$ mag at $\theta_{\rm Sun}<-8^\circ$, and
 $s_{br} >-0.4$ mag at $\theta_{\rm Sun}<-14^\circ$. 

\begin{figure}
\begin{center}
\epsfig{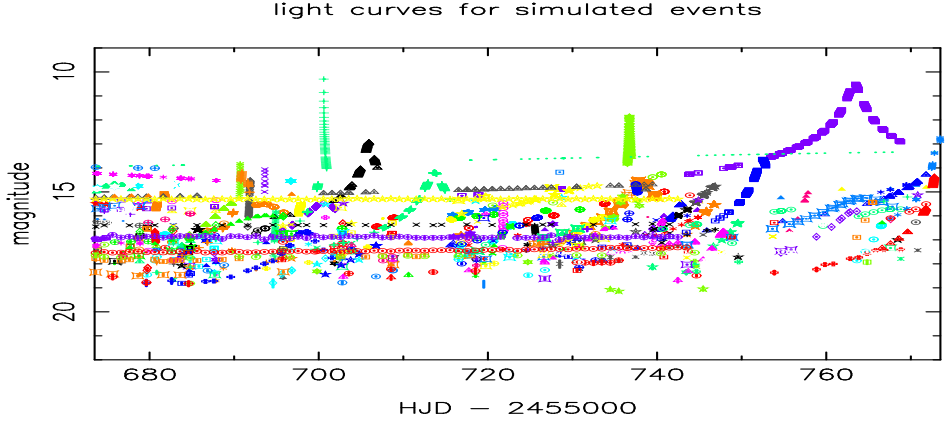}
 \caption{Light curves 
for events selected for OGLE observations. HJD is the Julian Date.} 
   \label{fig1}
\end{center}
\end{figure}

\begin{figure}
\begin{center}
\epsfig{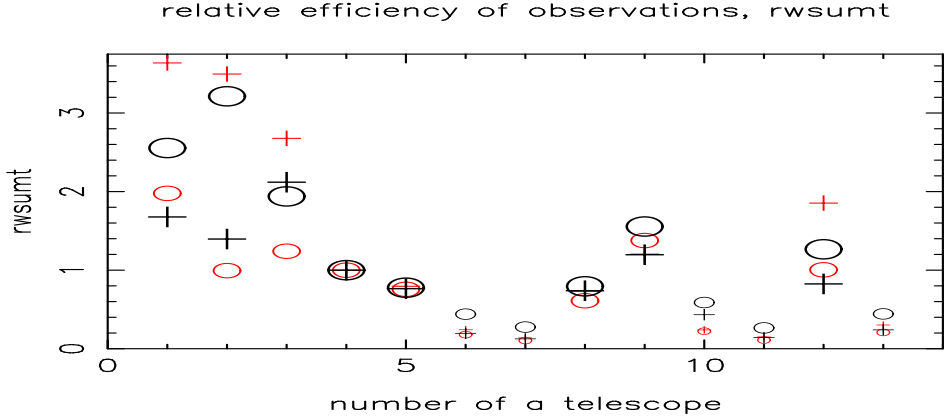} 
 \caption{The values of $r_{wsumt}$, which characterize the 
exoplanet detection capability of
observations (in the case of 1562 events available for observations),
 vs. the number $N_t$ of a telescope (see $N_t$ in Section 1) in the case when 1 m telescopes (equipped with the Sinistro CCD) located at the same site observe different events at the same time. 
Crests and ellipses
are for the 100-day time interval beginning from April 22 and August 1, 2011, respectively.
The signs for calculations with actual values of $t_0$ (the time corresponding to the peak of a light curve) 
and with random values of $t_0$ are black greater and red smaller, respectively. 
Small signs are for non-priority telescopes.
For random values of $t_0$, the number of light curve peaks was greater.}
   \label{fig2}
\end{center}
\end{figure}

{\underline {\it Selected events.}}
The code finds time intervals when it is possible to observe different events and the time intervals 
for `best' events (i.e., events selected for observations at a current time).
It also produces the plots of time variations of seeing, airmass, sky brightness, and flux for the best events.
Example of the light curves for the best events 
is presented in Fig. 1.
Below we analyze observations of 1562 OGLE events (110001-111562).
                                                                                    
{\underline {\it Exoplanet detection capability of observations.}}
Our simulator suggests what discovered events it is better to observe at specific time intervals with a specific telescope in order to maximize the exoplanet detection capability of observations.
To compare the capability for different telescopes, for the `best' events we considered the value of $w_{sum}$. The calculation of $w_{sum}$ is discussed in  
http://star-www.st-and.ac.uk/$\sim$si8/doha2013s.ppt.
In Fig. 2 for 13 telescopes, we present the values of 
$r_{wsumt}=(w_{sum}/w_{sum}^{OGLE})/(t_{sum}/t_{sum}^{OGLE}$), where $t_{sum}$ is the total time during considered time interval when it is possible to observe at least one event with a specific telescope, $t_{sum}^{OGLE}$ is the value for OGLE. 
The values of $r_{wsumt}$ (the capability relative to that for OGLE)
for the 2 m telescopes often are in the range of 1.4-2.1, 
but in some calculations they exceeded 3.
For the 1 m telescopes, $r_{wsumt}$ is often about 0.8, but in some runs it was greater (up to 1.6). 
The ratio of $w_{sum}$  for 1 m SSO with the Sinistro CCD to that for 2 m FTS located at the same site usually was about 1/2. 
For the SBIG CCD, the values of $w_{sum}$ were smaller by a factor of ~1.2 than those for the Sinistro CCD. 
The value of $w_{sum}$  was typically proportional to the diameter of the mirror.  
For 1562 events available for observations
and 100-day interval, a considerable (often $>$50\%) contribution to $w_{sum}$ was during short time
intervals corresponding to peaks of light curves if this telescope is allowed to observe all events.
The values of $r_{wsumt}$ can differ by a factor of up to 3 at a different choice of times $t_0$ corresponding to peaks of light curves.
If only events with the peak value of the magnification $A_{\max}$$>$$50$ (4\% of events) are observed,
then the ratio $r_{50}$ of the value of  $w_{sum}$ for such observations 
to the value of $w_{sum}$ for the observations allowed for all 1562 events can
differ by an order of magnitude for different considered time intervals.
For example, for OGLE   $r_{50}$ equals 0.07 and 0.99 for 5-day intervals starting from April 22 and August 1, 2011, respectively;
for 100-day intervals,  $r_{50}$ is 0.63 and 0.83, respectively.
For  $50$$<$$A_{\max}$$<$$200$ (2\% of events), the above values for  $r_{50-200}$ are 0.056, 0.78, 0.09, and 0.50, respectively.
Analysis of our calculations shows that during most of the time it is better to observe different events 
using different telescopes located at the same site than to observe the same event with two or three telescopes, 
but at the time close to a light curve peak often it is better to observe the same event with all telescopes located at the same site. 

\section{Conclusions}
     We have developed models for sky brightness and 
    seeing, calibrated by fitting to data from the OGLE survey and RoboNet observations in 2011. Time intervals during which events are observable are identified by accounting for positions of the Sun, the Moon and other  restrictions on telescope pointing. Simulated observations are then generated for an algorithm that adjusts target priorities in real time with the aim of maximizing planet detection zone area summed over all the available events.
Our code can be used for planning various observations (not only observations of microlensing events).
The obtained results show that, for a search for exoplanets based on already discovered events, 
a 2 m LCOGT telescope (FTS, FTN, or LT) is more effective (per unit of time of observations) than OGLE, 
and the efficiency of a 1 m telescope with the Sinistro CCD often is a about 0.8 of that of OGLE, but sometimes it can be  greater than that of OGLE.

This publication was made possible by NPRP grant
NPRP-09-476-1-78 from the Qatar National Research Fund (a member of Qatar
Foundation).  The statements made herein are solely the responsibility of
the authors.

\end{document}